\documentclass[12pt]{amsart}

\usepackage[english]{babel}
\usepackage{amsmath}
\usepackage{amsfonts}
\usepackage{amssymb}
\usepackage{amsthm}
\usepackage{latexsym}

\usepackage[dvips]{graphicx}
\usepackage{psfrag}
\title{Holomorphic Matrix Integrals}
\thanks{This paper is based on the ETH diploma thesis
of the second author}
\author{Giovanni Felder}
\author{Roman Riser}

\begin{document}

\maketitle
\centerline{ \it Department of Mathematics, ETH Zurich}
\centerline{ \it 8092 Zurich, Switzerland}

\begin{abstract}
We study a class of holomorphic matrix models.
The integrals are taken over middle dimensional 
cycles in the space of complex square matrices. As
the size of the matrices tends to infinity, the
distribution of eigenvalues is given by a measure
with support on a collection of arcs in the complex
planes. We show that the arcs are level sets of
the imaginary part of a hyperelliptic integral
connecting branch points.
\end{abstract}

\section{Introduction}

According to Dijkgraaf and Vafa \cite{Dij1,Dij3}, the effective glueball superpotential
of $\mathcal N=1$ supersymmetric $U(n)$ gauge theory has an asymptotic expansion given by the 
planar part of the topological 
expansion of a matrix model. To give the effective potential
for various vacua of the gauge theory, Dijkgraaf and
Vafa propose a formula obtained by saddle point expansion
of the matrix integral around different critical points. The
filling fractions, i.e., the fraction of the eigenvalues sitting
close to each of the critical points of the potential
are the parameters selecting the vacua.

In this paper we give a way to define non-perturbatively
(i.e., beyond the saddle point expansion) the matrix
integrals. Different integrals are integrals over
different cycles in the space of normal matrices. In
the case of generic polynomial potentials it is possible
to construct a cycle for each of the critical points, so
that all effective potentials considered by Dijkgraaf
and Vafa arise in the asymptotic expansion of our integrals. 
We note that the idea of integrating over eigenvalues along 
suitable contours in the complex plane appears in special
situations in \cite{D}. David's contours were also more recently considered
in \cite{La},  where a (non-holomorphic) modification of the
integral was proposed to obtain arbitrary filling factors.

A second result of this paper is the description of the
asymptotic distribution of eigenvalues for each critical 
point. The eigenvalues, as predicted by Dijkgraaf and Vafa,
lie asymptotically along arcs connecting 
branch points of a two-fold cover of 
the complex plane. We give
a reality condition which specifies the shape of the
arcs and the density of eigenvalues.
 
Before considering the case of general 
$N\times N$ matrices, it
is instructive to consider the case $N=1$ of ordinary integrals. 
Consider the Airy integral 
\[
u(x)=\int e^{\frac i\mu(xt+t^3/3)}dt.
\]
This integral gives a formal solution of the Airy differential equation
$-\mu^2u''(x)+xu(x)=0$. Indeed the saddle point expansion around
the two critical points of the integrand give the two linear
independent formal power series solutions. The non-perturbative
solutions can be obtained by integrating along paths in the
complex plane connecting any two of 
the three directions at infinity where the integrand goes to zero.
More generally,
 the integrals of the form $\int \exp(\frac i\mu(xt+P(t))dt$ 
for a generic polynomial $P(t)$ of degree $n$ are solutions of
a differential equation of order $n-1$. We show that a system of $n-1$ 
linearly independent solutions are obtained by choosing $n$
integration contours in the complex plane
going to infinity in directions where the integrand decays 
exponentially. Moreover the contours can be chosen
so that each of them passes through exactly one of the
$n$ critical points. It follows that the $n-1$ 
saddle point expansions at each of the critical points are 
the asymptotic expansions of true solutions given by convergent
integrals.

In the case of matrices we consider integrals of
the form 
\begin{equation}\label{e-hmi}
\langle F\rangle_N=
Z_N^{-1}\int_\Gamma F(\Phi)\exp\left(-\frac N\mu
\,\mathrm{tr}\,p(\Phi)\right) d\Phi
\end{equation}
over real $N^2$-dimensional cycles $\Gamma$ in
the space of complex $N\times N$ matrices.
The potential $p$ is a polynomial with complex
coefficients,
 $d\Phi=\wedge_{j,k} d\Phi_{jk}$ and
the normalization factor $Z_N$ is such that
$\langle 1\rangle_N=1$. The observables
$F(\Phi)$ are holomorphic functions
invariant under conjugation.
 In the well-studied
case of integrals over hermitian matrices, with a polynomial $p$ with real coefficients  bounded below, the
saddle point approximation becomes exact in the large
$N$ limit with fixed $\mu$ and the relevant critical
point is described by an asymptotic
 {\em density of eigenvalues} which has
support  on a union of intervals on the real axis. If
$\mu$ is small, these intervals are small neighborhoods
of the minima of the potential $p$. 
To obtain the densities of eigenvalues needed to make
contact to gauge theory one considers the variational
problem for critical points subject to the side condition
that the fraction of eigenvalues in the vicinity of
each of the critical points (not just minima) are given
numbers.

We consider here the case of a generic polynomial
$p$ of degree $n$
with complex coefficients, with distinct critical
points $z_1,\dots,z_{n-1}$, and propose to consider
integrals over cycles in the space of normal matrices
(a matrix is normal if it commutes with its adjoint or,
equivalently, if it is conjugated to a diagonal matrix
by a unitary matrix). The cycles we consider are 
parametrized by integers $N_1,\dots,N_{n-1}$ whose
sum is $N$ and are characterized by the condition
that $N_k$ eigenvalues belong to a path $\Gamma_k$
in the complex plane
going through $z_k$ and going to infinity in a direction
where $\mathrm{Re}(p(z))\to\infty$. In the
limit $N\to \infty$ with $n_k=N_k/N$ fixed, the
eigenvalues in the saddle point approximation (supposed
to be exact in the limit) are distributed along
$n-1$ arcs in the complex plane. For small $\mu$ the
arcs are close to the critical points and the
$k$th arcs contains the fraction $n_k$ of the
eigenvalues.

\section{The one-dimensional case}
Let $p(z)=a_nz^n+\cdots+a_1z+a_0$ be a 
polynomial of degree $n\geq 2$ with
complex coefficients. 
We want to consider integrals of the form
\[
\int_\Gamma q(z)e^{\textstyle{-\frac1\mu p(z)}}dz,\qquad\mu>0.
\]
for polynomials $q(z)$.
Before considering the question of integration cycles
we may evaluate such integrals as asymptotic
series as $\mu\to 0$ by formal application of the
saddle point method at each of the critical points
$z_1,\dots,z_{n-1}$, which we assume to be
distinct.
In this way we get $n-1$ asymptotic series
of the form $\exp(-p(z_k)/\mu)(c_1+c_2\mu+\cdots)$
and the question is whether these are asymptotic
expansions of our integral for suitable cycles 
$\Gamma_k$.

The cycles which we should consider here are
(linear combinations of) paths for which the
integral converges. As the integrand is holomorphic,
homotopic paths will give the same answer and what
matters is the behavior at infinity. As $z\to\infty$,
$p(z)\sim a_n z^n$, so there are $n$ directions
in the complex plane for which 
$\mathrm{Re}(p(z))\to +\infty$ as $z$
tends to infinity in these directions. Let us
call these asymptotic directions valleys as in these
directions the integrand decays exponentially.
Neighboring valleys are separated by hills, which
are directions of exponential increase of the
integrand.
So the cycles one needs to consider are linear
combinations of infinite
paths connecting  pairs of distinct valleys. As paths
connecting two valleys can be deformed into sums
of paths connecting the two valleys with any third
one, there are only $n-1$ linearly independent
cycles.

Let us assume for simplicity that the critical
values $p(z_k)$ have distinct imaginary parts.
Then there is a canonical way to associate to
each critical point $z_k$ a path $\Gamma_k$
in such a way that the asymptotic expansion
of the integral over $\Gamma_k$ as $\mu\to 0$
is obtained by the saddle point expansion
at $z_k$. Namely, we take the {\em steepest descent} paths(see, e.g., \cite{bru} and \cite{erd})
emerging from $z_k$, defined by the
condition that the tangent vector at each
point points in the direction of the
gradient of $\mathrm{Re}(p(z))$. As $p(z)$
is holomorphic, the gradient of $\mathrm{Re}\,p(z)$
is orthogonal to the gradient of $\mathrm{Im}\,p(z)$
by the Cauchy--Riemann equations. Thus steepest
descent paths are level lines for the imaginary
part of $p(z)$. Each non-degenerate
critical point $z_k$ is at the intersection of 
two such level lines. One of these two lines, the
one along which $\mathrm{Re}(p(z))$ takes its minimum
at $z_k$, is the steepest descent path $\Gamma_k$.
Along the other line $\Gamma_k'$, the real part
takes its maximum at $z_k$.
We claim that $\Gamma_k$ is a smooth path 
going to infinity in both directions and connecting
two valleys. Indeed, $\Gamma_k$ is (in suitable
parametrization) given by a solution of the
differential equation $\dot z(t)=\overline{p'(z(t))}$.
It follows that $\frac d{dt}\mathrm{Re}(p(z(t)))=
|p'(z(t))|^2$. By our assumption, critical
values have distinct imaginary parts, so the
steepest descent path passing through $z_k$
may not come close to any other critical point.
Thus $|p'(z(t))|$ is bounded below so that, 
as we go away from $z_k$,
$\mathrm{Re}\,p(z(t)$ must go to infinity and the
path $\Gamma_k$ connects two valleys. 
Similarly $\Gamma'_k$ connects two hills. As
$\Gamma_k$ and $\Gamma'_k$ cross at $z_k$, the
two valleys connected by $\Gamma_k$ are separated
by hills and are thus different.

We have thus shown that for each critical
point $z_k$ there is a  steepest
descent path $\Gamma_k$ going through $z_k$
and connecting pairs of different valleys.
On $\Gamma_k$ the real part of $p(z)$ is minimal
at $z_k$ so that the saddle point expansion
at $z_k$ indeed gives the asymptotic expansion
of the integral over $\Gamma_k$.

\section{Matrix integrals}
\subsection{Integration cycles}
We consider matrix integrals of the form \eqref{e-hmi}
for $p(z)$ a polynomial of degree $n$.
They are integrals of holomorphic differential forms 
over $N^2$ dimensional cycles. For each set of
natural numbers $N_1,\dots, N_{n-1}$ summing up to $N$,
we have a cycle in the normal matrices, characterized
by the condition that $N_k$ eigenvalues run over
the path $\Gamma_k$ of the previous section.
More precisely, the cycle $\Gamma$ is parametrized
by $\mathbb R^N\otimes U(N)/U(1)^N$:
\[
(t,U)\mapsto U\,
\left(\begin{array}{ccc}
\lambda_1(t_1)&    & \\
        &\ddots & \\
        &      & \lambda_N(t_{N})
      \end{array}
\right)
\,U^{-1}.
\]
The first $N_1$ diagonal elements
 $\lambda_1(t),\dots, \lambda_{N_1}(t)$ are parametrizations of
$\Gamma_1$, the next $N_2$ are parametrizations of $\Gamma_2$
and so on.

The usual argument to reduce the integral to an integral
over the eigenvalues (see \cite{BIZ}) gives
\[
\langle F\rangle_N=\frac 1{Z'_N}
\int F(\lambda_1,\dots,\lambda_N)
e^{\textstyle{-\frac 
N\mu\sum_{j=1}^n p(\lambda_j)}}\prod_{j<k}(\lambda_j-\lambda_k)^2
d\lambda_1\cdots d\lambda_N,
\]
The integral is over $\Gamma_1^{N_1}\times\cdots\times\Gamma_{n-1}
^{N_{n-1}}$ and the function $F$, a function on matrices
invariant under conjugation, is regarded as a
symmetric function of the eigenvalues.

\subsection{The loop equation}
To study the large $N$ limit, 
it is useful to introduce the trace of the
resolvent: 
\[
\omega(z)=\frac1N\,\mathrm{tr}\left(\frac1{z-\Phi}
\right)
         =\frac1N\sum_{j=1}^N\frac1{z-\lambda_j},
\]
as products of such traces are generating functions of 
polynomial functions invariant under conjugation.
The ``loop equation'' \cite{W,M} for this quantity is
\[
\mu\langle\omega(z)^2\rangle_N-p'(z)\langle
\omega(z)\rangle_N+f_N(z)=0,
\]
where 
\[
f_N(z)=\frac1N
\left\langle \sum_i\frac{p'(z)-p'(\lambda_i)}{z-\lambda_i}
\right\rangle_N
\]
is a polynomial of degree $n-2$ with leading 
coefficient $na_n$.
This equation can be derived from the identity
\[
0=\sum_{i=1}^N\int\frac\partial{\partial\lambda_i}
\left(\frac1{z-\lambda_i}e^{-\frac N\mu\sum_j
p(\lambda_j)}\prod_{j>k}(\lambda_j-\lambda_k)^2
\right)\prod d\lambda_j.
\]
In the limit $N\to\infty$, the matrix integral
is supposed to be dominated by an integral
over a region where the eigenvalues are close
(for small $\mu$)
to critical points of $p$. With our choice of
integration cycles and keeping $n_j=N_j/N$
fixed as $N\to\infty$, there will be a fraction
$n_j$  of the eigenvalues close to $z_j$.

In the limit $N\to\infty$ one expects that 
the saddle point approximation becomes exact
and thus  $\langle\omega(z)\rangle_N$ converges to 
\[
\langle\omega(z)\rangle=\int \frac1{z-z'}d\nu(z'),
\]
for some probability measure $\nu$ with support
in regions around the critical points of $p$ and
so that $n_k$ is 
the measure of the region around $z_k$.
Technically one assumes that the limit $N\to\infty$
exists and that $\lim_{N\to\infty}
(\langle\omega(z)^2\rangle_N-\langle\omega(z)\rangle_N^2)
=0$. 
The function $\langle\omega(z)\rangle$ is defined and
holomorphic for $z$ outside the support of the measure.
Setting 
\begin{equation}\label{e-y}
y(z)=2\mu\langle\omega(z)\rangle-p'(z),
\end{equation}
we finally obtain
\begin{equation}\label{e-hy}
y(z)^2=p'(z)^2-4\mu f(z),
\end{equation}
for some polynomial $f(z)$ of degree $n-2$ with
leading coefficient $na_n$. Thus the function $y(z)$,
which is a priori defined on the complement of the
support of the measure, has an analytic continuation
to a two-fold covering of the complex plane. The
original function $y$ is the branch of the function
defined by \eqref{e-hy} which behaves at infinity
as $-p'(z)$.

\subsection{The density of eigenvalues}
With the analogy with the case of hermitian matrices in
mind, it is reasonable to assume that the measure
$\nu$ has support on a collection of
arcs $\gamma_1,\dots,\gamma_{n-1}$ which for small
$\mu$ are close to the critical points of $p$, and
that $d\nu(t)=\rho(t)dt$ for some {\em density 
of eigenvalues} $\rho$ defined on
the arcs: namely, the measure $\nu(U)$ of a
set $U$ intersecting one of the arcs, say $\gamma_j$,
in a piece $\gamma_j\cap U$ is
\[
\nu(U)=\int_{\gamma_j\cap U}\rho(z)dz.
\]
Note that for the right-hand side to be defined we
need to fix an orientation on $\gamma_j$.
  Then $y(z)$ is a holomorphic function
outside on the complement of the arcs and the density
on $\gamma_j$ is related to the discontinuity of $y$:
\[
\rho(x)=\frac1{4\pi i\mu}(y(x^-)-y(x^+)),\qquad x\in\gamma_j.
\]
Here $y(x^+)$ ($y(x^-)$)
 denotes the limit of $y(z)$ as $z$
tends to $x$ from the left (from the right)
 of the oriented curve $\gamma_j$.

{}From this information we deduce that the arcs 
$\gamma_j$ connect pairs of zeros of $y$,
the branch points of the hyperelliptic
curve $y^2=p'(z)^2-4\mu f(z)$ of genus $n-2$. 
The measure
of the $j$th arc is then the period
\begin{equation}\label{e-pe}
n_j=\frac1{4\pi i\mu}\int_{A_j}y(z) dz ,\qquad j=1,\dots, n-1,
\end{equation}
over a cycle $A_j$ enclosing the pair of branch
points in counterclockwise direction. It follows from
the condition on the leading coefficient of $f$ that
$\sum n_j=1$.
We also note that
 since $y(x^-)=-y(x^+)$, we have the formula
$\rho(x)=y(x^-)/2\pi i\mu$.

\subsection{The shape of the arcs}
There remains to determine the precise form
of the arcs and the coefficients in $f$ as functions
of the filling fractions $n_j$.
First of all, the relation between the $n-2$ 
free complex coefficients $b_0,\dots, b_{n-3}$ of $f(z)$
(recall that
the leading coefficient is fixed to be $b_{n-2}=
na_n$) and the periods $n_j$ ($1\leq j\leq n-1$)
subject to $\sum n_j=1$ is, locally around any
point where the branch points are distinct,
 a holomorphic diffeomorphism,
since the Jacobian matrix
\[
\frac{\partial n_j}{\partial b_{k-1}}=-\frac 1{4\pi i\mu}
\int_{A_j}\frac {z^{k-1}}ydz, \qquad j,k=1,\dots,n-2.
\]
is non-degenerate, being the matrix of $a$-periods
of a basis of Abelian differentials on a smooth
curve. It follows that
locally there exists a real $(n-1)$-dimensional submanifold
in the complex space $\mathbb C^n$ of coefficients $b_j$
which maps to real positive $n_j$.

The condition that fixes the shape of the arcs is
the reality and positivity condition for the
density: if $t\mapsto z(t)$ is a parametrization
of the arc $\gamma_j$ respecting its orientation,
the condition is
\begin{equation}\label{e-re}
\rho(z(t))\frac {dz(t)}{dt}\geq 0.
\end{equation}
Using $\rho(z)=y(z^-)/2\pi i\mu$, we may parametrize
the arcs (away from the endpoints)
to be solutions of the differential equation
\[\dot z(t)=i\,\overline{ y(z(t)^-)}
,\]
connecting branch points. Alternatively, arcs
may be described as level lines of a function: 
introduce the hyperelliptic integral
\[
F(z)=\frac 1{4\pi i\mu}\int_{z_0}^z y(z) dz.
\]
It is a holomorphic many-valued function on the complement
of the support of the measure. As we go around 
an arc $\gamma_j$, $F(z)$ increases by $n_j$ so
$\mathrm{Im}\,F(z)$ is single valued.
The measure of a piece between
two points $x_1$, $x_2$ on a curve $\gamma_j$ is  $F(x_2^-)-
F(x_1^-)$ which is real.
Thus the arcs $\gamma_j$ are level lines of the imaginary
part of $F(z)$. 
Around a branch point $z_0$ which is a simple
zero of $y^2$,
$F(z)\sim \mathrm{const}+ (z-z_0)^{3/2}$. Therefore
there are three smooth 
level lines of $F$ emerging from every
simple branch points. In the most general
situation the support of the measure may then be
a graph consisting of level lines of $\mathrm{Im}\,F$
joining branch points. 
A more precise description is possible in the case
of small $\mu$ to which we turn.

\subsection{Small coupling}
\begin{figure}[h]
\begin{picture}(150,130)(60,0)\scalebox{0.55}
{\includegraphics{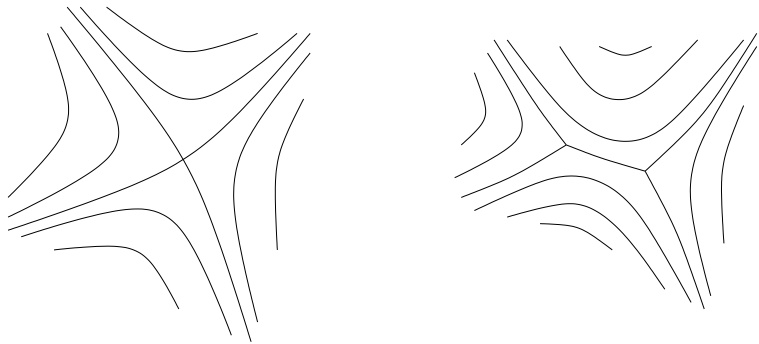}}
\end{picture}
\caption{Level lines of $\mathrm{Im}\,F$ for $\mu=0$
(left) and for $\mu>0$ (right)}\label{f-1}
\end{figure}

For small $\mu>0$ we claim that the arcs and the
density of eigenvalues are determined completely
by the filling fractions $n_j$ through \eqref{e-hy}
and \eqref{e-re}. 
To show this, notice first that as $\mu\to 0$,
pairs of branch points $z_j',z_j''$ converge
to the critical points $z_j$ and the periods $n_j$
(eq.~\eqref{e-pe}),
regarded as functions of $\mu$ and the coefficients
$b_j$ of $f$ are holomorphic at $\mu=0$. We have
\[
\lim_{\mu\to 0} n_j(\mu,b_0,\dots,b_{n-3})=
\frac{f(z_j)}{4\,p''(z_j)}.
\]
Since there is a bijective holomorphic
correspondence between
values $f(z_j)$ at the distinct points $z_j$ and
coefficients $b_j$ of $f$, we have at $\mu=0$,
and by analyticity also for small $\mu$, a
biholomorphic map $(b_0,\dots,b_{n-3})\to(n_1,\dots,n_{n-1})$,
 $\sum n_j=1$. In particular, we can invert this
map and find a unique $f$ for each set of $n_j\geq0$,
such that $\sum n_j=1$.
It remains to show that for all small $\mu>0$ there
is a level line $\gamma_j$ of $\mathrm{Im}\,F(z)$
connecting $z_j'$ to $z_j''$. As $\mu\to 0$,
$\mu F(z)\to 
-\frac1{4\pi i}p(z)+\mathrm{const}$. The level lines
of $\mu\,\mathrm{Im}\,F(z)$ at $\mu=0$ 
are thus the level lines of $\mathrm{Re}\,p(z)$.
In the neighborhood of a non-degenerate
critical point $z_j$ they
look like the left picture in Fig.~\ref{f-1}: on any 
small circle around $z_j$ each value is taken on at
most four times. 
For positive small $\mu$ the critical point $z_j$ splits 
into two branch points $z_j'$, $z_j''$ from
each of which three level lines emerge. The
condition that the period $n_j$ is real implies that 
$\mathrm{Im}\,F(z_j')=\mathrm{Im}\,F(z_j'')$. The
function $g(z)=\mathrm{Im}\,(F(z)-F(z_j'))$ is defined
up to a sign in a neighborhood
of $z_j',z_j''$, so that its zero level line
 is uniquely defined. It
follows that there is a level line $\gamma_j$ of $\mathrm{Im}\,F$, namely the zero set of $g$, joining $z_j'$
to $z_j''$ as on the right picture in Fig.~\ref{f-1}:
if none of the level lines emerging from $z_j'$ and
$z_j''$ were to join, $g(z)$ would take
the value zero six times on any small circle 
encircling $z_j',z_j''$, which cannot be, as this
does not happen at $\mu=0$. Also, a level line cannot go
from a point $z_j'$ or $z_j''$ to itself as the 
real part of $F$ is monotonic along level lines.

We conclude that for any small $\mu>0$, and any
given filling fractions $n_1,\dots,n_{n-1}\geq 0$
summing up to 1, there is a unique polynomial
$f(z)=na_nz^{n-2}+b_{n-3}z^{n-3}+\dots+b_0$, so that
the curve $y^2=p'(z)-4\mu f(z)$ has $a$-periods $n_j$.
The zeros of $y$ are connected in pairs by arcs
$\gamma_j$ obeying the reality condition \eqref{e-re}.
According to the discussion above, these arcs are
are the support of the measure and the density of
eigenvalue is $(2\pi i\mu)^{-1}y(x^-)$, $x\in\gamma_j$.

\section{Concluding remarks}
We have given a non-perturbative definition of 
the matrix integrals that in the large $N$ limit
give the superpotentials considered in \cite{Dij1,Dij3}.
We considered the case of a generic polynomial
potential with complex coefficients. For
small 't Hooft coupling $\mu$, the density
of eigenvalues was shown to be given by arcs
connecting pairs of branch points of a
hyperelliptic curve. The shape of the arcs is
uniquely determined by a reality condition.
For larger $\mu$ or for potentials with degenerate
critical points, one expect the arcs to combine into
graphs in the complex plane. It would be interesting
to understand what kind of graphs can arise in this
way.

\end{document}